# OPTICAL AND ELECTRONIC PROPERTIES IN FERROELECTRIC BARIUM TITANATE-BASED COMPOUNDS

Halyna Volkova[1,*], Pascale Gemeiner[1], Grégory Geneste[2], Jérôme Guillot[3], Carlos Frontera[4], Nidal Banja[1], Fabienne Karolak[1], Christine Bogicevic[1], Brahim Dkhil[1], Nicolas Chauvin[5], Damien Lenoble[3], and Ingrid C. Infante[5]

[1] Laboratoire Structures, Propriétés et Modélisation des Solides, CentraleSupélec, CNRS-UMR8580, Université Paris-Saclay
8-10 rue Joliot-Curie, Gif-sur-Yvette, France

[2] CEA, DAM
DIF, Arpajon, France

[3] Luxembourg Institute of Science and Technology, Materials Research and Technology Department
41 rue du Brill, Belvaux, Luxembourg

[4] Institut de Ciència de Materials de Barcelona, Consejo Superior de Investigaciones Científicas
Campus de la UAB, Bellaterra, Spain

[5] Institut de Nanotechnologies de Lyon, CNRS-UMR5270 ECL INSA UCBL CPE, Université de Lyon
7 avenue Jean Capelle, Villeurbanne, France

**Abstract.** *The bandgap energy values for the ferroelectric $BaTiO_3$-based solid solutions with isovalent substitution $Ba_{1-x}Sr_xTiO_3$, $BaZr_xTi_{1-x}O_3$ and $BaSn_xTi_{1-x}O_3$ were determined using diffuse reflectance spectra. While the corresponding unit cell volume follows Vegard's law in accordance with the different ionic radii of the ionic substitutions, the bandgap values depict non-linear compositional dependences for all the solid solutions. The effect is considerably large for $BaZr_xTi_{1-x}O_3$ and $BaSn_xTi_{1-x}O_3$ solutions, depicting a bandgap linear compositional dependence up to x=0.6, for x>0.6 $BaZr_xTi_{1-x}O_3$ compounds present much larger bandgap values than $BaSn_xTi_{1-x}O_3$ counterparts. Electronic properties have been investigated through X-ray photoelectron spectroscopy in $BaSn_xTi_{1-x}O_3$ compounds, indicating that the Sn 3d and Ti 2p core levels shift against the Ba 3d ones within the whole compositional range with the same energy trend as that observed for the optical bandgap. Since for $Ba_{1-x}Sr_xTiO_3$ compounds no major bandgap variation is observed, we conclude that the bandgap compositional dependences observed for $BaSn_xTi_{1-x}O_3$ compounds and $BaZr_xTi_{1-x}O_3$ ones are originated from the structural sensitivity of the O, Ti and Sn or Zr electronic bands involved in the bandgap transition of these compounds. With this work, we underline the reliability of the bandgap determined from diffuse reflectance spectrometry experiments, as a means to non-invasively evaluate the electronic properties of powder materials.*

## 1 Introduction

Since the theoretical efficiency limit of single p-n junction solar cell was calculated by Shockley and Queisser [1], there is a challenge of finding new ways to bypass it. A way to intrinsically increase the efficiency could be to use alternative materials such as ferroelectrics (FEs). Unlike p-n junctions, the potential difference in ferroelectrics arises from their non-centrosymmetric unit cell providing the so-called bulk photovoltaic effect [2]. Among them, $BaTiO_3$ is a well-referenced, relatively cheap to produce, and environmentally friendly FE. In view of finding new ways to obtain FE-control of the optical properties, $BaTiO_3$-solid solutions have been studied, as potentially depicting FE-order with simultaneously enhanced photoconductive properties. Here we present the work on three solid solutions with isovalent substitutions, in view of understanding the pure structural and electronic

---

*Corresponding author: H. Volkova (halyna.volkova@centralesupelec.fr)

H. Volkova, P. Gemeiner, G. Geneste, J. Guillot, C. Frontera, N. Banja, F. Karolak, C. Bogicevic, B. Dkhil, N. Chauvin, D. Lenoble, and I. C. Infante effects and minimizing the point defect contributions. With $Sr^{2+}$ substituting $Ba^{2+}$, the $Ba_{1-x}Sr_xTiO_3$ is a family of solid solutions with ferroelectricity remaining at low temperature up to x_Sr~0.9 [3]. $BaZr_xTi_{1-x}O_3$ and $BaSn_xTi_{1-x}O_3$ are solid solutions with $Ti^{4+}$ being substituted by $Zr^{4+}$ and $Sn^{4+}$, respectively, thus presenting conduction band where Zr 4d and Sn 5s 5p states will be added to Ti 3d ones, which in principle will depict much different effective mass and mobility, for the potential photoconduction properties that we are seeking. The phase diagrams of $BaZr_xTi_{1-x}O_3$ [4] and $BaSn_xTi_{1-x}O_3$ [5] display rich polar features, evolving at room temperature from a pure tetragonal FE in the Ti-rich region towards a paraelectric cubic structure in the Ti-poor region. In-between, the compositions are FE-relaxors characterized by polar disorder and average cubic structure.

## 2 Experimental procedures

$Ba_{1-x}Sr_xTiO_3$ (x_Sr = 0, 0.25, 0.5, 1), $BaZr_xTi_{1-x}O_3$ (x_Zr = 0, 0.1, 0.2, 0.3, 0.4, 0.5, 0.6, 0.7, 0.8, 0.9, 1) and $BaSn_xTi_{1-x}O_3$ (x_Sn = 0, 0.1, 0.2, 0.4, 0.5, 0.6, 0.8, 0.9, 1) samples were synthesized by solid state reaction from starting materials, $BaCO_3$ (99.9%), $SrCO_3$ (99.9%), $TiO_2$ (99%), $ZrO_2$ (99.9%), and $SnO_2$ (99.9%). After homogenization in a mortar, by ultrasonic bath and by magnetic mixing, the powders were calcined at temperatures between 800°C and 1200°C, and subsequently ground with polyvinyl alcohol and pressed into pellets. Two annealing steps at 600°C and 800°C insured the evaporation of binding agent. The sintering was held at temperatures of 1280-1450°C adjusting the dwell times, resulting in pellet density of ~92-95%. Structural evaluation by X-ray diffraction (XRD) was performed using a Bruker D2 Phaser diffractometer on finely ground and annealed powders, with 0.02º step and acquisition times adjusted for different angular ranges (10 sec/step for 2θ=20-35°, 30 sec/step for 35-60°, 55 sec/step for 60-120°). Unit cell volume was determined using Le Bail analysis through devoted crystallographic software (Jana2006). The ultraviolet-visible-near-infrared spectrometry experiments were performed on a Perkin Elmer Lambda 850 spectrometer in diffuse reflectance geometry using a Harrick's Praying Mantis™ accessory, from finely grinded and annealed powders to minimize mechanical stresses. From the raw reflectance, $R$, the absorption coefficient $k$ is determined according to Kubelka-Munk reemission function $F_{KM}$ transformation [6], within the assumption the scattering coefficient $s$ is independent of the photon energy:

$$F_{KM} = \frac{k}{s} = \frac{(1-R)^2}{2R}. \qquad (1)$$

The optical bandgap is determined from a linear fit to the onset of the absorption edge obtained after converting the $F_{KM}$ following Tauc formalism, which for the present study we consider as direct bandgap. X-ray photoelectron spectroscopy (XPS) studies were carried out on powder and ceramic samples. Axis Ultra DLD spectrometer with Al Kα X-ray source operated at 50W was used, together with charge neutralizer. For the analysis of the XPS data (CasaXPS), we have used the internal reference of the Ba $3d_{5/2}$ core level energy, fixed at 778.5eV, as considered to be the ion with very limited changes in its electronic state due to chemical, electronic and structural effects.

## 3 Results and Discussion

From the XRD collected (Figure 1a-c), we deduce the stabilization of a pure perovskite phase for each solid solution compound, with unit cell volume following Vegard's law (Figure 1d). The larger compositional slope for Zr-compounds than for Sn-ones is in agreement with the expected effect induced by the difference in ionic radius of $Zr^{4+}$ against $Sn^{4+}$, depicting the $Sr^{2+}$-compounds a volume reduction, as expected [8].

Optical properties of the different compositions are shown in Figure 2a-d. Diffuse reflectance raw data in Figure 2a point out the large absorption differences of parent compounds. Optical absorption features are seen in the Kubelka-Munk functions $F_{KM}$ depending of the solid solution. A large increase of $F_{KM}$ is characteristic of the optical absorption edge at a given photon energy, being the

H. Volkova, P. Gemeiner, G. Geneste, J. Guillot, C. Frontera, N. Banja, F. Karolak, C. Bogicevic, B. Dkhil, N. Chauvin, D. Lenoble, and I. C. Infante

Zr-compounds (Figure 2c) those presenting the larger variation of the photon energy onset. No remarkable onset change is seen for Sr-compounds (Figure 2b), and weak and non-monotonic ones for Sn-compounds (Figure 2d). The chemical substitution effect on the occupied electronic states is directly investigated on Sn-based compounds through XPS core level acquisition. Using Ba3d$_{5/2}$ state as the perovskite energy reference, we plot in Figure 2e-k the different characteristic O1s, Sn3d and Ti2p occupied states. Fine analysis of these levels indicate a shift of the energy positions of Sn and Ti as a function of the Sn-content (Figure 2i and k).

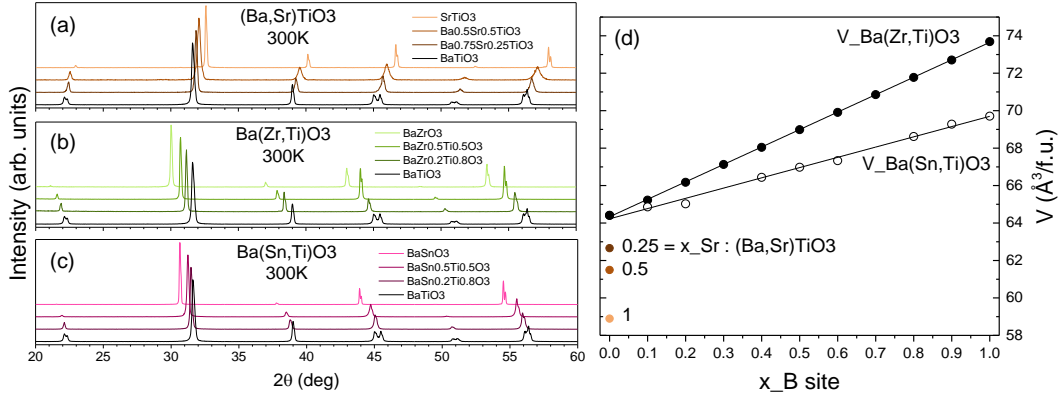

Figure 1: Room temperature X-ray diffraction [(a)-(c)] powder patterns and (d) unit cell volume determined from the XRD patterns, for different BaTiO$_3$-based compounds. [(a)-(c)]: (a) Ba$_{1-x}$Sr$_x$TiO$_3$ with x_Sr=0, 0.25, 0.5, 1 (bottom to top), (b) BaZr$_x$Ti$_{1-x}$O$_3$ with x_Zr=0, 0.2, 0.5, 1 (bottom to top), and (c) BaSn$_x$Ti$_{1-x}$O$_3$ with x_Sn=0, 0.2, 0.5, 1 (bottom to top); (d) Pseudocubic cell volume per formula unit as a function of x_B site (Zr or Sn) (symbols), corresponding linear fits are shown.

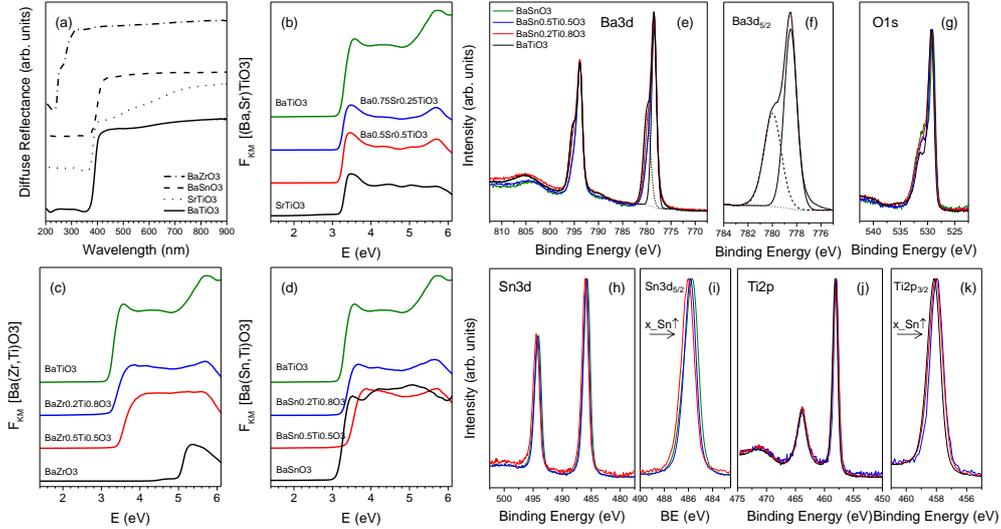

Figure 2. [(a)-(d)] Optical properties from BaTiO$_3$-solid solutions, and [(e)-(k)] X-ray photoelectron spectroscopy (XPS) core level data from Ba(Sn,Ti)O$_3$. [(a)-(d)]: (a) Diffuse reflectance spectra from parent compounds (BaTiO$_3$, SrTiO$_3$, BaSnO$_3$ and BaZrO$_3$, bottom to top), and [(b)-(d)] Kubelka-Munk FKM functions vs photon energy E determined for (b) Ba$_{1-x}$Sr$_x$TiO$_3$ with x_Sr=0, 0.25, 0.5, 1 (top to bottom), (c) BaZr$_x$Ti$_{1-x}$O$_3$ with x_Zr=0, 0.2, 0.5, 1 (top to bottom), and (d) BaSn$_x$Ti$_{1-x}$O$_3$ with x_Sn=0, 0.2, 0.5, 1 (top to bottom). [(e)-(k)]: Core levels from BaSn$_x$Ti$_{1-x}$O$_3$ compounds (x_Sn=0, 0.2, 0.5, 1) (a) Ba3d, (g) O1s, (h) Sn3d, (i) Sn3d$_{5/2}$, (j) Ti2p, (j) Ti2p$_{3/2}$, together with the example on BaTiO$_3$ sample of the different components used for Ba3d$_{5/2}$ fitting (Ba within BaCO$_3$, dash line, Ba within the perovskite -here, BaTiO3- fill line, and background, dotted line).

H. Volkova, P. Gemeiner, G. Geneste, J. Guillot, C. Frontera, N. Banja, F. Karolak, C. Bogicevic, B. Dkhil, N. Chauvin, D. Lenoble, and I. C. Infante

Bandgap values as a function of Sn- or Zr- substitution (Figure 3a) account for the large differences previously noticed from direct analysis of the optical properties through $F_{KM}$. Remarkably, for x_B > 0.2, both solution compounds depict polar disorder, being relaxors at room temperature, follow the same initial trend up to x_B ~ 0.6. For x > 0.6, Zr-compounds present larger bandgap values while lower ones are determined for Sn-compounds. Comparing the obtained optical bandgap values for Sn-compounds with the energy differences of the XPS Ti2p and Sn3d electronic bands (Figure 3b-c), we notice a similar substitution dependence. Moreover, these XPS energy differences are fully in agreement with the observed bandgap variation, being for optical and XPS results this energy difference of ~ 0.35eV between $BaTiO_3$ and x_Sn = 0.8. Investigations are ongoing to determine the precise origin of this behavior and its coupling to the relaxor properties.

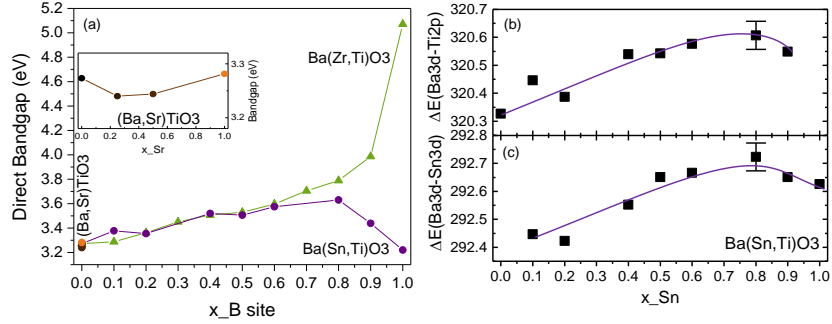

Figure 3: (a) Bandgap values as a function of the x_B site substitution for different $BaTiO_3$-solid solutions, determined using $F_{KM}$ and Tauc plot linear fits, assuming direct bandgap. Inset: Bandgap for the $Ba_{1-x}Sr_xTiO_3$ compounds *vs* x_Sr substitution. [(b),(c)]: B-site core level energy difference (ΔE) relative to the perovskite $Ba3d_{5/2}$ core level position for different $BaSn_xTi_{1-x}O_3$ compounds and as a function of the x_Sn substitution, being (b) $Ti2p_{3/2}$ and (c) $Sn3d_{5/2}$ (lines are a guide for the eye).

## 4 Conclusions

Combining structural, optical and electronic characterization tools on pure perovskite $Ba_{1-x}Sr_xTiO_3$, $BaZr_xTi_{1-x}O_3$ and $BaSn_xTi_{1-x}O_3$ solid solutions, we proved that we can precisely determine their bandgap values. We have successfully presented the potential of diffuse reflectance spectrometry for the determination of the optical and electronic properties of single phase perovskite oxides, and similar approaches can be exploited for investigating the optical transitions of other materials.

## REFERENCES

[1] W. Shockley and H. J. Queisser. Detailed Balance Limit of Efficiency of p-n Junction Solar Cells, *Journal of Applied Physics*, 32, 510. 1961.

[2] A. M. Glass et al. High-voltage bulk photovoltaic effect and the photorefractive process in $LiNbO_3$, *Applied Physics Letters*, 25, 233. 1974.

[3] C. Ménoret et al. Structural evolution and polar order in $Sr_{1-x}Ba_xTiO_3$, *Physical Review B*, 65, 224104. 2002.

[4] T. Maiti, R.Guo and A. S. Bhalla. Structure-Property Phase Diagram of $BaZr_xTi_{1-x}O_3$ System, *Journal of the American Ceramics Society*, 91, 1769. 2008.

[5] X. Wei and X. Yao. Preparation, structure and dielectric property of barium stannate titanate ceramics, *Materials Science and Engineering: B*, 137, 184. 2007.

[6] B. Philips-Invernizzi, D. Dupont and C. Cazé. Bibliographical review for reflectance of diffusing media, *Optical engineering*, 40, 1082. 2001.

[7] J. Tauc, R. Grigorovici, and A. Vancu. Optical Properties and Electronic Structure of Ge, *Physica Status Solidi*, 15, 627. 1966.

[8] R. D. Shannon. Effective Ionic Radii in Oxides and Fluorides, *Acta Crystallographica*, B25, 925. 1969.